\documentclass{optica-article}
\usepackage{svg}
\usepackage{algpseudocode}
\usepackage{float}
\journal{opticajournal}
\articletype{Research Article}
\usepackage{xurl}
\begin{document}

\title{Effective Training Principles of Physical Reservoirs}


\author{Sobhi Saeed$^{1,\dagger}$, Mehmet Müftüoglu$^{1,\dagger}$,Glitta R. Cheeran$^{1}$, Juliane Heim$^{1}$, Bennet Fischer$^{1}$, and Mario Chemnitz$^{1,2,*}$
}
\vspace{+2em}

\author{$^{\dagger}$ These authors contributed equally to this work \\
    $^{1}$ Leibniz-Institute of Photonic Technology, Albert-Einstein-Str. 9, 07745 Jena, Germany \\
    $^{2}$ Institute of Applied Optics and Biophysics, Friedrich Schiller University Jena, Helmholtzweg 4, 07745 Jena, Germany \\
    $^{*}$ Corresponding author, mario.chemnitz@leibniz-ipht.de}

\vspace{+3em}  

\begin{abstract*}
 Reservoir computers benefit from the inherent complexity of optical phenomena, which provide rich, often nonlinear dynamics. However, training directly on the reservoir's output renders the system prone to overfitting and computationally ineffecient during the training phase. In this work, we investigate strategies to mitigate overfitting and reduce computational overhead through output pruning and regularization. We compare loss-minimizing search methods (Equal Search and Branch $\&$ Bound) against an output-oriented statistical filtering approach (Variance Filter) and random pruning, highlighting advantages and disadvantages of each approach and the overall importance of informed reservoir output sampling, particularly for a shrinking latent space. We further demonstrate that enforcing readouts selection across the full output spectrum improves performance, especially for non-iterative methods. Additionally, we examine L1 and L2 regularization techniques (LASSO and ridge regression), both of which significantly enhance performance on highly nonlinear tasks such as the Spiral Benchmark. While our methods are of general use, results are obtained from and discussed exemplary for a nonlinear fiber-optical extreme learning machine. Overall, this study provides a deep analysis of reservoir hidden layer filtering mechanisms and output layer training, enabling optimized performance in physical reservoir computing systems.
\end{abstract*}\newline
\vspace{+3em}  

\newpage 
\section{Introduction}
Reservoir computing was proposed to be unified by Verstraten et. al. in 2006 \cite{rc_proposal} to gather similar network structures, which are specific types of networks in the category of recurrent neural networks (RNNs), under a common framework despite having different names. Echo state networks \cite{echostate_lukosevicius_reservoir_2009} and Liquid State Machines \cite{Liquid_State_Maas} are the foundational models of reservoir computing (RC). The RC is a specific type of RNN which has three layers, i.e. input layer, hidden layer or reservoir and the output layer. The input layer and reservoir has randomly initialized weights and biases and only output weights which connects reservoir layer to output layer are trainable. The RC's training scheme is linear regression to perform prediction or classification tasks \cite{rc_review_tanaka_recent_2019}. The Extreme Learning Machine (ELM) is alternative version of RC by eliminating the fading memory or recurrency of the RC.
The advantage of RC and ELM is their computationally light training scheme, which consists of linear regression on the response of the reservoir\cite{elm_origin_aritcle_HUANG2006489}. Since the idea is not to tweak the internal weights of the reservoir and only the output weights are trained using linear regression, the reservoir response becomes the main remaining component to optimize. 
There are several strategies to alter the response of the reservoir. These include tuning the operating regime of the reservoir by controlling the lyapunov exponent \cite{RC_lyapunov_control}, tuning the fading memory which is increasing the weight of self connecting neurons in the reservoir \cite{RC_fading_memory_tuning}, filtering the readout layer \cite{Pruning_in_RC_dutoit_pruning_2009}, introducing inputs with intrinsic embeddings of the reservoir \cite{RC_embeddings}, and adding regularization terms to the regression \cite{RC-ridge_regression}.

Reservoir computing and extreme learning machines have been demonstrated on different platforms. Scattering media \cite{ScatteringRC_pierangeli_photonic_2021}, microresonators \cite{RC-Massar_integrated}, semiconductor laser networks \cite{rcwithrealandvirtualnodes_rohm_multiplexed_2018}, VCSELs with diffractive optics \cite{brunner_reconfigurable_2015}, and highly nonlinear fibers \cite{ELM_Bennet_HNLF} are among the implementation platforms in the optical domain. Most of these platforms do not provide adequate results on the first attempt. Therefore, physical reservoirs or ELMs require tuning. This tuning can be performed on several aspects, including the nonlinearity, the number of readout layer nodes, the type of readout layer nodes (for reservoirs, virtual nodes are sometimes preferred), and the method for how input is introduced. The common requirement among all these aspects is the definition of a metric and a parameter space. These are the main elements of any optimization problem. Therefore, a formal optimization problem definition for parameter tuning would provide a solid basis for solving the tuning problem. This tuning can be separated into two categories: changing the dynamics of the reservoir or hidden layer, and changing the implementation of the model without altering the dynamics. Optimizing the Lyapunov exponent, the number of principal components, and the fading memory of the reservoir belong to the first category. Changing the regularization term in the regression, optimizing the number of readout nodes, and modifying the input of the reservoir belong to the second category.

There have been systematic studies on both categories. In category one; lyapunov exponent is an metric that quantifies how much the nonlinear system diverge or converge from the current state. Moreover, controlling this metric to increase the accuracy of the given task is demonstrated \cite{RC_lyapunov_control}. The principle component analysis shows how many essential dimensions in the given dataset \cite{pca_Matilde}. The inference capacity of an physically realized ELM is estimated with this analysis \cite{RC_PCA}. In category two; pruning the reservoir/hidden layer in ELMs is discussed \cite{Pruning_in_RC_dutoit_pruning_2009} , regularization methods for regression are compared \cite{CHEN_norms}, the combination of regularization and pruning is demonstrated \cite{dutoit_pruning_2009}, and variance filtering is presented \cite{RC-Filtering}. Each approach has advantages and disadvantages. Pruning either takes more time while finding the optimal solution, or takes less time at the cost of forfeiting optimality. Regularization overcomes outliers, noise, and overfitting, but adds computational cost and increases execution time. Variance filtering offers fast calculation time and a task-agnostic approach, but it does not provide an optimal solution. 

In this paper, we selected nonlinear fiber as the platform for the physical reservoir/hidden layer because supercontinuum generation provides complex dynamics in the spectral domain \cite{ELM_Bennet_HNLF}. Moreover, we implemented an Extreme Learning Machine approach since our platform does not exhibit fading memory. We specifically focused on solving classification problems, the MNIST handwritten digits dataset, and the spiral benchamrk \cite{Saeed}. While MNIST is a high-dimensional problem with moderately nonlinear decision boundaries, the spiral benchmark is low-dimensional but requires highly nonlinear decision boundaries. We approached these problems through the second tuning category, namely pruning or sampling the readout layer and regularizing the regression algorithm. Firstly, we formulated the scheme as a well-known optimization problem (01-Knapsack) \cite{Knapsack_Mathews}. Secondly, we approached the problem using two optimization algorithms: Equal Search and Branch$\&$Bound \cite{Land_BB}. Thirdly, we introduce an output-oriented search method, Variance Filter with more advanced criteria, we further investigate Random Pruning. Finally, we regularized the regression using two different regularization methods.

\section{Methods}
\subsection{Experimental Setup}
All data used in this work have been acquired with a fiber-optical Extrem Learning Machine (FO-ELM, see Fig. 1a) that comprises a femtosecond laser source (Toptica DFC) operating at 80 MHz repetition rate with 70 nm bandwidth centered at 1,560 nm, followed by an erbium-doped fiber amplifier (Thorlabs EDFA300P) operated at 25$\%$ pump current to compensate for system losses. Data are encoded using a polarization-maintaining programmable spectral filter (Coherent Waveshaper 1000A/X) operating in the extended C- to L-band (1,528–1,602 nm), which allows to independently modify spectral phase and amplitude across 400 frequency channels for both dispersion compensation and phase information imprinting. Placing the Waveshaper after the EDFA enables precise controlling of the attenuation, and, thus, the input power to the nonlinear fiber circumventing gain saturation effects and nonlinear deterioration of the encoded phase information before being processed.\newline
The initially chirped pulses are precompressed through a dispersion-compensating fiber (Thorlabs PMDCF, 1~m length) to 400~fs before entering the spectral filter. A particle swarm optimization algorithm in conjunction with optical autocorrelation measurements has been utilized to further optimize the input pulse phase to obtain 135~fs \cite{Bennet_spie, Efimov_98} before entering the main processing unit--a highly nonlinear fiber. Two nonlinear fibers were in use: (1) Thorlabs HN1550, 5~m length with all-normal dispersion ($-1$~ps/nm/km at 1,550~nm), and 10.8~W$^{-1}$km$^{-1}$ nonlinear coefficient, and (2) Thorlabs PMHN1, 5~m length, with anomalous dispersion ($+1$~ps/nm/km at 1,550~nm), and 10.7~W$^{-1}$km$^{-1}$ nonlinear coefficient. The output is measured using an optical spectral analyzer (Yokogawa AQ6375E) and analyzed on a standard office computer. The data encoding on the optical pulse is described in the previous study \cite{Saeed}. We used the normal dispersion fiber for all experiments on MNIST and the anomalous dispersion fiber for the spiral benchmark. Training was performed on the output spectra, where each output is represented as a one-dimensional vector of intensity values (readouts) sampled across discrete wavelength bins. The concatenation of these vectors forms the reservoir state matrix $X$.
\begin{figure}[H]
    \centering
    \includegraphics[width=12cm]{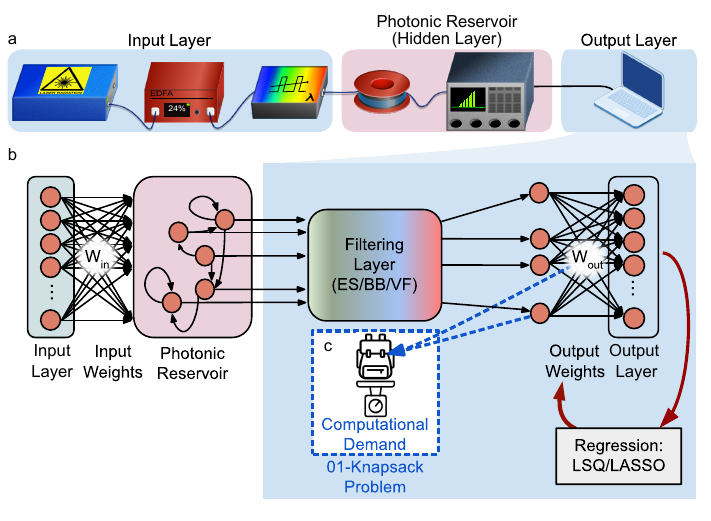}
    \caption{Physical reservoir computing diagram. a) Experimental realization of physical reservoir computing. The input layer consists of a femtosecond laser, and Erbium-doped amplifier (EDFA) and a spectral encoder. The hidden layer is Highly Nonlinear Fiber (HNLF) and a Optical Spectrum Analzyzer samples the spectrum. The output layer takes the spectral samples and does the training of reservoir computing. b) Physical reservoir computing flow-chart. Input layer connection introduces data to hidden layer with input weights $W_{in}$ as randomly assigned. Photonic reservoir process information with linear or nonlinear information mixing. Hidden Layer Filtering selects the most essential readout dimensions with Equal Search (ES) or Branch$\&$Bound (BB) or Variance Filtering (VF) algorithms. The output weights $W_{out}$ connects the filtered output of photonic reservoir to the output layer with the help of least ordinary squares (LSQ) or least absolute shrinkage and selection operator (LASSO) regression techniques. The output layer indicates the predicted label of the task. The training scheme conssists of Hidden Layer Filtering and regression trainning with the output layer values. c) 01-Knapsack problem diagram. The selection of nodes with filtering layer and output weights are the selection parameters to maximize the accuracy.}
    \label{fig:placeholder}
\end{figure}
\subsection{Knapsack Problem}
\textbf{The model of reservoir computing in the principle has the advantage of utilizing physical systems' diverse dynamics with a closed form solution of least ordinary squares (LSQ) \begin{equation}
W = (X^TX + \lambda I)^{-1}X^Ty
\end{equation}
to predict complex functions such as time series prediction, regression, and classification.} In the physical implementation especially in photonics there are many dimensional outputs from the photonic reservoir. The overfitting problem is unavoidable at that point. Therefore, in the literature there are several approaches to tackle this problem. Decreasing the readout layer’s dimensions (pruning) and implementing regularizations to regression are proposed solutions to solve the high dimensionality problem. All of those proposed solutions are actually solving an optimization problem. In the idea, while regularization effectuates the most effective readout dimensions to maximize the accuracy in the task, pruning finds out least important readout dimensions and disconnects them from the output layer to maximize accuracy.\newline
Both approaches find an optimal solution inside the provided solution space. Therefore, the formulization of this problem into a well known discrete optimization problem paves the way to select solutions and compare solutions in a systematic way. Thereafter, we formalized the pruning and weight matrix calculation of a physical reservoir computer as an 01-Knapsack problem. There is a clear analogy between the 01-Knapsack problem and the classification task with physical reservoirs. Each class is the knapsack, each readout dimension is the items that you fit inside the knapsacks, and the weights are the location and orientation of the items. At this point, the problem becomes; is there an optimal way to put all necessary items into bins in an ideal orientation such that it has the least weight. The weight in this analogy is the mean square error (MSE) between the prediction and the labels. After constructing this analogy we structured the problem as in below. 
Let m denote the number of prediction targets, n the total number of wavelengths, L is the label vector, and K the number of spectral shots. We define a binary decision variable
\[x_j \in \{0,1\}, \quad j = 1,\ldots,n,\]
where $x_j = 1$ indicates that wavelength $j$ is selected and $x_j = 0$ otherwise.
The prediction for target $i$ is defined as a linear combination of the
selected wavelengths:
\[p_i(\mathbf{x}) = \sum_{j=1}^{n} w_{ij} x_j,\]
where $w_{ij}$ represents the contribution of wavelength $j$ to prediction $i$.
The weight $w_{ij}$ is decomposed into spectral-shot and prediction-specific
terms as
\[w_{ij} = \sum_{k=1}^{K} w_{ik} \, I_{kj},\]
where $I_{kj}$ denotes the intensity value of wavelength $j$ in spectral
shot $k$, and $w_{ik}$ is the weight of spectral shot $k$ for prediction $i$.
The objective is to minimize the total squared prediction error (Loss):
\[\mathcal{L}=\min_{\mathbf{x}} \; \sum_{i=1}^{m} \left( p_i(\mathbf{x}) - L_i \right)^2.\]
The wavelength selection is constrained by a cardinality (knapsack) constraint:
\[\sum_{j=1}^{n} x_j \leq c,\]
where $c$ denotes the maximum number of wavelengths that may be selected.
The resulting 01-knapsack optimization problem is therefore open to tackle with optimization algorithms comparable on the same problem domain.

\subsection{Hidden Layer Filtering}
RC can be utilized for the classification tasks as proposed in \cite{RC_on_image_classificaiton_jalalvand_real-time_2015} even though the linear regression is not the ideal algorithm to solve classification. Therefore, in this study we extended linear regression’s output with a decision condition in which the winner takes it all. Among the given continuous results the highest valued vector is selected as the predicted class \cite{Saeed}.
In this section, we use MNIST handwriten digits calssification task to compare Hidden Layer Filtering algorithms. The comparison’s main metric is accuracy (defined in this context as the ratio of the correctly classified data to the full data) which is obtained after executing linear regression, and then "winner takes it all" on the spectral output. In addition, we also compare the achieved accuracy while operating on the raw data (MNIST input images) to highlight the effect of nonlinear mapping of the physical reservoir, following the same procedure for the algorithms that have reasonable execution time. We use the first 2400 images to train and test the system with a 4:1 split ratio, 192 training images per class. The images consist of gray pixels with integer values from 0 to 255. Before encoding the images on the spectral phase of optical pulses (photonic reservoir), we flatten and normalize them to values between 0 and 1 before multipling with a scaling factor. The waveshaper encodes the data on the spectral phase. An output spectrum corresponding to each individual image is recorded with an OSA between 1380 nm and 1800 nm with 0.2 nm resolution. In the rest of the script for the Hidden Layer Filtering method, we used this spectra dataset to compare each pruning algorithm’s performance..
\subsubsection{Equal Search}
In this part, we are going to introduce the Equal Search(ES) algorithm and its performance on the data. This algorithm can be considered in the approximate/Exact algorithm category \cite{ELM_Bennet_HNLF}.
In the Equal Search, the goal is finding out the optimal combination of spectral bins that is sampled for MNIST task and normalized pixels. This algorithm takes a set of pixels or spectral bins, number of desired bins or pixels, and loss function as input. Then it initiates a loss matrix which has the desired number of bins as rows and mode of the input set on the desired number of bins as columns. Afterwards, it starts a nested loop with the first loop scans the starting point of the selection grid and the second loop sweeps the separation of this grid. In each step of the loop, the algorithm assigns the loss value to the loss matrix. The minimum valued indice is selected as the optimal solution. Finally, the found solution is converted into a set of spectral bins and the algorithm returns an optimal spectral bin set or pixel set with the size of desired number of elements as output.\newline
The comparison between pixels of MNIST images and spectrally encoded MNIST images are demonstrated in Fig. 2. In Fig. 2a, accuracies of spectral bin and image pixel search over different numbers of set size plot is demonstrated. It can be observed from the Fig. 2a that spectral bin search does not have an overfitting effect between training and testing as strong as in the pixels. While ES on spectral bins plateau around 90$\%$ test accuracy, ES on pixels plateau 80$\%$. This shows that reservoir is essential to improve the capability of linear regression. The Figs. 2b and 2c show where the selected bins are located on the MNIST image and how the loss matrix looks. From Fig. 2b, the optimal solutions are distributed in equal radius around the center of the image. The Figs. 2d and 2e show the same specifics of the algorithm on spectral bins. There is a gradual convergence around the encoding bin region in the spectra when the number of bins increases towards 100. In the 10 bins case the ratio between bins are located at the encoding region and the rest is increasing parallelly as the number of bins. Finally, for the Handwritten MNIST task ES achieved 91$\%$ accuracy with 300number of bins.
The main advantage of this algorithm is that it is relatively fast among the “trail and error” search algorithms because it has a strong parsimony condition to meet. This condition decreases the number of possible solutions from the total solution space enormously. However, this relative speed advantage is the positive outcome of trading the optimality. The main disadvantage of this algorithm is no guarantee of the optimal solution and no estimation to optimal solution.

\begin{figure}[H]
    \centering
    \includegraphics[width=8cm]{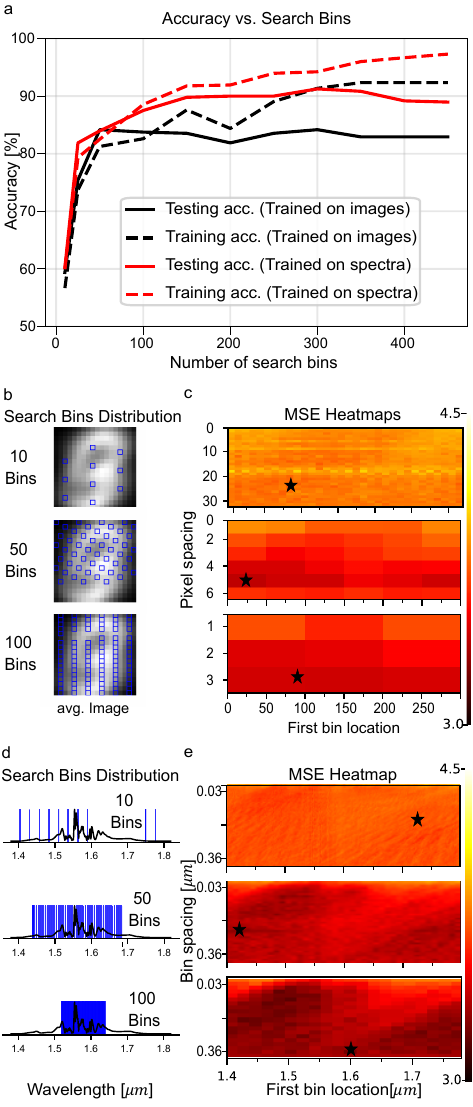}
    \caption{Equal Search (ES) performance comparison on Handwritten MNIST images
classification. (a) Testing and training accuracies vs. number of search bins (readouts) when training on input images vs. output spectra. (b) Bins selected by ES (training on images), shown on the average image for 10, 50, and 100 bins. (c) Corresponding loss heatmaps indicating where the loss is lowest; loss values are indicated by the attached colorbars. The x-axis indicates the starting bins, and the y-axis indicates the pixel spacing, the star shows the start and bin spacing of the best combination of equally distanced bins which yielded the lowest loss, shown in b. (d) Bins selected by ES (training on spectra), shown on the average spectrum for 10, 50, and 100 bins. (e) Corresponding loss heatmaps indicating bins selection when training on the ouput spectra.}
    \label{fig:placeholder}
\end{figure}

\subsubsection{Branch$\&$Bound}
In this part, we are going to introduce the Branch$\&$Bound(BB) algorithm and its performance on the data. This algorithm can be considered in the approximate algorithm category.
The algorithm minimizes the loss value while subjecting to find a set with the desired number of bins. This algorithm takes a set of pixels or spectral bins, probability of the bins or pixels, capacity ratio to limit the search space, loss function, and the decrease size as input. It calculates the maximum number of branches from the total number of given bins and the decreased size. Thereafter, it generates a candidate set that includes equals to or less than capacities ratio of possible solutions with the decreased number of bins. It gets their loss value and uses the lowest loss valued 1000 candidates to update the probability of the decreased number of bins. Then the best subset is selected and the parameters are updated. It continues until the maximum number of branches is reached.\newline
In the BB implementation to the physical reservoir the selected solutions seem distributed along all wavelengths in the first look but some regions have more density than the other regions. Likewise, the encoding region (between 1530 and 1600) is the most dense area in the spectrum. This algorithm’s main advantage is the fact where the optimal solution would be. The pruning ratio and searched space ratio are equal to each other, the optimal solution observed. This relation’s prediction of optimal solutions shown in Fig. 3c. For the case of searching 10$\%$ of the total solution space and with the pruning step 15 bins indicate the optimal solution at 165 bins. Moreover, we observed this prediction is fitting to our execution of the algorithm. The minimum loss is obtained at 165 bins in Fig. 3. However, one of the main disadvantages is continuing the pruning doesn’t suffice the minimization. The other disadvantage is that the necessary number of iterations to reach the optimal solution. 

\begin{figure}[H]
    \centering
    \includegraphics[width=14cm]{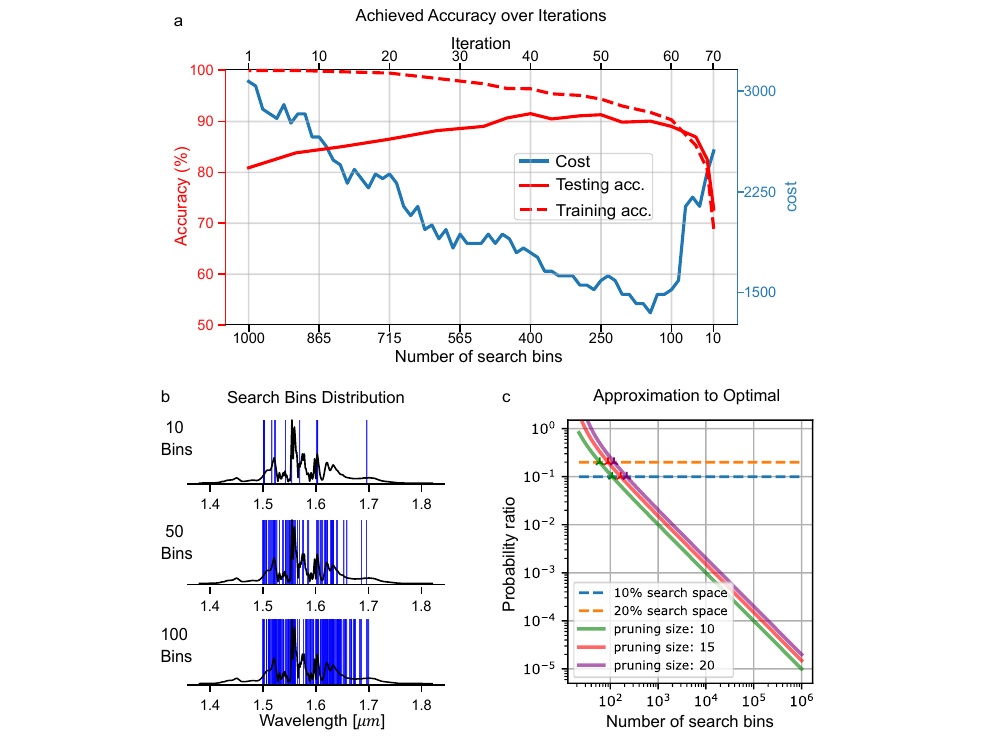}
    \caption{Illustration of Branch$\&$Bound (BB) algorithm on Handwritten MNIST images classification. (a) Testing and training accuracies and cost over iterations with pruning size is 15 in each step. (b) Bins selected by BB, shown on the average spectrum for 10, 50, and 100 bins. (c) Approximaiton of pruning with respect to the search space. The dashed lines are 10$\%$ (blue) and 20$\%$ (orange) probabilities. The solid lines are the ratio of size of search space after pruninginitial to size of the search space. While the crosses are located at 110, 165, and 220 number of search bins for 10$\%$, the other crosses are located at 60, 90, and 120 number of seach bins in order to pruning size 10, 15, and 20.}
    \label{fig:placeholder}
\end{figure}

\subsubsection{Variance Filter}
In this section, we introduce an alternative approach for filtering the latent search space. The proposed method selects search bins based on their standard deviation (SD) across all training output spectra (or input images, when training is performed on images), independent of class labels. This corresponds to a well-established statistical technique known as the Variance Filter (VF) \cite{bommert_multicriteria_2017}.\newline
Within the Extreme Learning Machine (ELM) framework; where only the output layer is trained, selecting bins with the highest SD effectively identifies the most sensitive spectral regions. These regions represent “windows” in which the readouts exhibit the greatest variation in response to changes in the encoding signal. As a result, this method enables sensitivity-aware pruning of the search space, similar to the approach described in \cite{jafari_sensitivity-guided_2026}.
A key characteristic of this method is its output-oriented nature. Rather than relying on iterative trial-and-error procedures to determine informative bins, it directly applies statistical analysis on the spectral outputs, leading to a more efficient and principled selection process.\newline
Our initial approach involved selecting $N_{global-max}$ bins corresponding to the global maximum SD. However, we observed that these selected search bins tended to cluster around the central region of the spectrum, neglecting other regions of the latent space, as illustrated in Fig. 4f while training on spectra. A similar pattern could also be detected while training on images (see Fig. 4c). This central concentration limits the system’s performance in both cases, as demonstrated in Fig. 4a. This clustering is attributed not only to higher responsivity in the central region but also to the generally higher intensity around the central pump region.\newline
To address this bias, we employed normalized standard deviation (NSD), defined as the SD divided by the average intensity. While bins with the highest NSD values still appeared mostly at the central pump region, they also emerged in other spectral regions, providing a more balanced bin distribution, yet still neglecting most regions of the supercontinuum as shown in Fig. 4g.\newline
This approach yields a comparable performance when the system is trained on output spectra (see Fig. 4a); however, it presents several limitations when training on input images. Specifically, most pixels with high NSD are observed near the edges due to very low average intensity in these sections (see Fig. 4d), leading to a poor selection of search bins in these low-information areas, resulting in bad performance as illustrated in Fig. 4a.\newline
To enforce bin selection spans the entire spectral range, we implemented a local search strategy. This approach partitions the full spectrum into smaller intervals and identifies the local maximum SD within each interval. The interval length depends on the ratio between the total available bins (controlled by the recording resolution) and the user-specified number of search bins (e.g., 50).\newline
This local search strategy ensures a balanced distribution of bins across all spectral regions when training on output spectra, and across all spatial areas when training on images as shown in Fig. 4e and 4b, respectively. By preventing the over-representation of any single region, this approach mitigates overfitting and promotes more robust model generalization, yielding in better performance overall (see Fig. 4a).
\begin{equation*}
\begin{aligned}
l &= \frac{N_{\text{total}}}{N_{\text{local}}} \\
 N_{\text{total}} &= \frac{wl}{r}
\end{aligned}
\end{equation*}
$N_{\text{total}}$ is the total number of points per record, $l$ is length of the local interval, $N_{\text{local}}$ is the desired number of local search bins (user choice), $r$ is the spectral resolution (nm), and $wl$ is the wavelength interval (nm).
\begin{figure}[H]
    \centering
    \includegraphics[width=14cm]{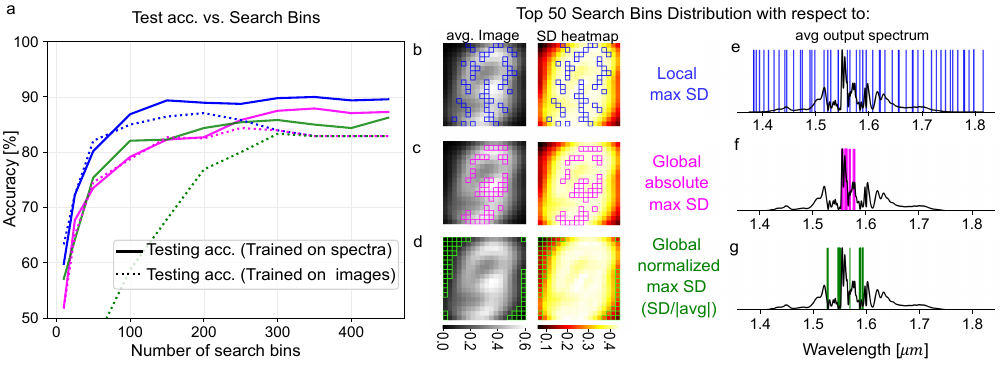}
    \caption{Comparison of different Variance Filter (VF) search approaches (testing on Handwritten MNIST images classification). (a) Testing accuracies vs. number of search bins when training on input images and output spectra, using three different search approaches: Local Maximum Search (LMS, blue), Global Maximum Search (GMS, magenta), and Global Normalized Maximum Search (GNMS, green). (b–d) Bins selected by VF\_LMS, VF\_GMS, VF\_GNMS when training on input images. (e–g) Bins selected by VF\_LMS, VF\_GMS, VF\_GNMS when training on spectra.}
    \label{fig:placeholder}
\end{figure}
After achieving promising results using the three search strategies, we were motivated to explore whether a combined approach could further improve classification performance. The idea is to leverage the strengths of each method by allocating a proportion of the total number of bins $(N_{\text{total}})$ to each method.
To implement this, we define a splitting ratio that determines how many bins are selected based on each method. For example, a ratio of (0.1–0.1–0.8) corresponds to:
\begin{equation*}
\begin{aligned}
N_{\text{global-max}} &= 0.1 \times N_{\text{total}} \\
N_{\text{global-norm-max}} &= 0.1 \times N_{\text{total}} \\
N_{\text{local}} &= 0.8 \times N_{\text{total}}
\end{aligned}
\end{equation*}
$N_{\text{global-max}}$ is the number of bins corresponding to global absolute maximum SD, $N_{\text{global-norm-max}}$ is the number of bins corresponding to global normalized maximum SD, and $N_{\text{local}}$ the number of bins corresponding to local maximum SD. \newline
Before combining the selected bins, we ensure that no duplicates exist across the three subsets to avoid redundancy. Using this combined method with a (0.2–0.0–0.8) split, we achieved classification accuracies of 92.3\% on the MNIST dataset when training on spectra and 86\% when training on input images, both using 200 search bins (Fig. 5a). The newly selected search bins appear in Figs. 5(b–g) for bin counts of 10, 50, and 100 under both training approaches.\newline
We attribute the improved performance to better class clustering achieved by combining global and local searches, as shown in Fig. 7h. For class "3", the spectral intensity distribution (averaged over selected bins only) centered tightly around the overall average (blue dashed line) when using local search, causing 20.5\% misclassification. Using global absolute maximum search shifted the distribution rightward (magenta dashed line) and widened it excessively, resulting in 25.6\% misclassification. The combined approach (shown in purple) produced tighter, well-bounded clustering, reducing misclassification to 15.4\%. Other classes, such as "1" and "7", showed less benefit since their corresponding spectral intensity outputs already exhibited strong clustering regardless of search method.\newline
Although determining the optimal split ratio requires empirical tuning, repeated testing across benchmarks consistently favored configurations where 80\% or more of the bins originated from local search, with the remainder from the two global searches. Variance Filter mostly outperforms a well-established search method like Equal Search while requiring significantly less runtime (cf. Figs. 7g–i and table 2)

\begin{figure}[H]
    \centering
    \includegraphics[width=14cm]{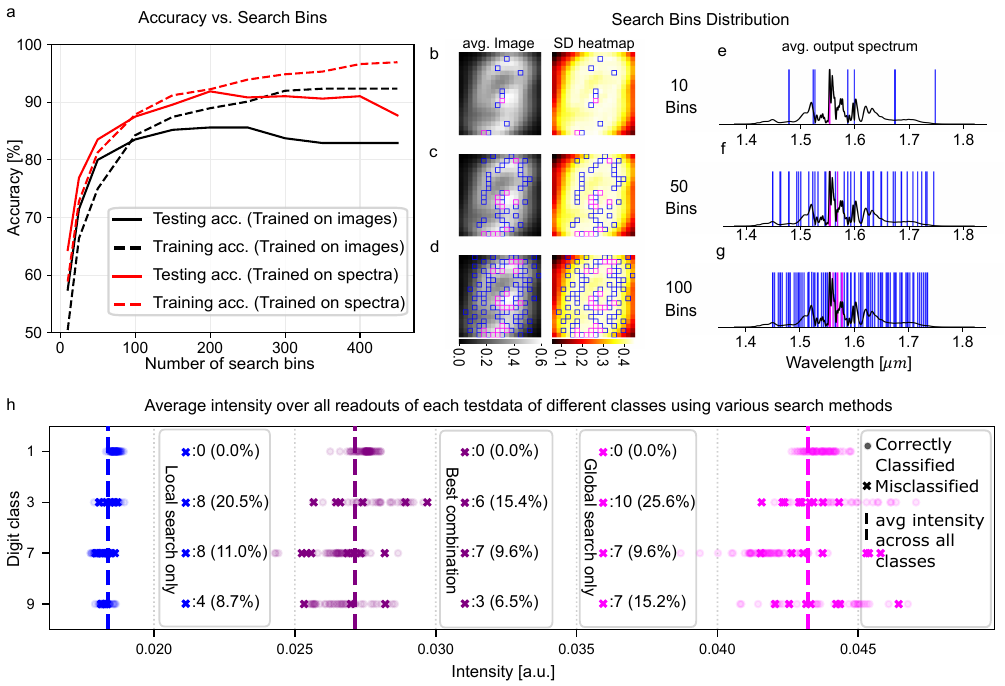}
    \caption{Performance of combined Variance Filter (VF) search approaches for handwritten MNIST classification. (a) Testing and training accuracy versus number of search bins for input images and output spectra. (b–d) Bins selected by the optimal VF combination: Local Maximum Search (LMS, 80\%, blue) and Global Absolute Maximum Search (GAMS, 20\%, magenta) at 10, 50, and 100 bins (input images). (e–g) Corresponding selections when trained on output spectra. (h) Intensity distributions across test data classes comparing LMS only (blue), GAMS only (magenta), and combined approach (purple; 80\% LMS + 20\% GAMS), illustrating class clustering improvements..}
    \label{fig:placeholder}
\end{figure}

\subsubsection{Random Search}
Random feature selection has long been explored as an efficient strategy to develop scalable search methods, notably exemplified by random forests \cite{dudek_comprehensive_2022}, where a random subset of the latent space is selected to reduce computational complexity while maintaining representational capacity. Such approaches can also be applied to sample the outputs of physical reservoir computers \cite{zheng_random_2025}.\newline
Extreme learning machines (ELM) leverage random initialization of weights and connections \cite{elm_origin_aritcle_HUANG2006489}. Furthermore, subsequent studies have investigated randomly pruning inactive hidden nodes in ELM architectures \cite{mahmood_svmelm_2016}, as well as performing targeted pruning of irrelevant input variables \cite{yoan_miche_op-elm_2010}, demonstrating that stochastic selection mechanisms remain effective across diverse neural computing paradigms.\newline
In this section, we evaluate a fully randomized bin selection approach applied on the ELM output layer and compare its performance against the optimized search methods introduced previously. The comparison is based on training and testing accuracy achieved on the MNIST benchmark, with models trained directly on both input images and spectral data representations.\newline
Fig. 7a demonstrates that randomly selecting search bins yields performance comparable to algorithmically optimized bin selection. This efficacy stems from an intrinsic property of Random Search (RS): it generates a quasi-symmetric distribution that ensures adequate spatial coverage across the latent space. As established in the preceding section, forcing bin selection to span the entire search domain enhances performance; RS inherently satisfies this constraint due to its stochastic nature, thereby providing sufficient representational diversity without explicit optimization.\newline
However, random selection exhibits notable instability. As illustrated in Fig. 7c, test accuracy varies substantially depending on the random seed, especially when the system is trained on input images even while using high number of search bins. While such configurations (lucky seeds) cannot be directly searched for, one could optimize in the vicinity of promising random sampling and treat them as initialization points for further refinement toward the optimal solution—a search approach known as semi-randomized selection \cite{wang_novel_2020}.\newline
Another critical limitation of random search emerges when the search interval shrinks. As demonstrated in Fig. 7b, which compares the achieved testing accuracy using 100 search bins selected by RS, Equal Search (ES) and Variance Filter (VF) while progressively narrowing the spectral search interval. RS exhibits significant performance degradation regardless of the selected random seed compared to the optimized pruning techniques (ES and VF).\newline
We hypothesize that this behavior arises, because identifying informative readouts becomes increasingly harder as the latent space shrinks, highlighting that optimized pruning becomes essential when the available feature space is confined.

\begin{figure}[H]
    \centering
    \includegraphics[width=14cm]{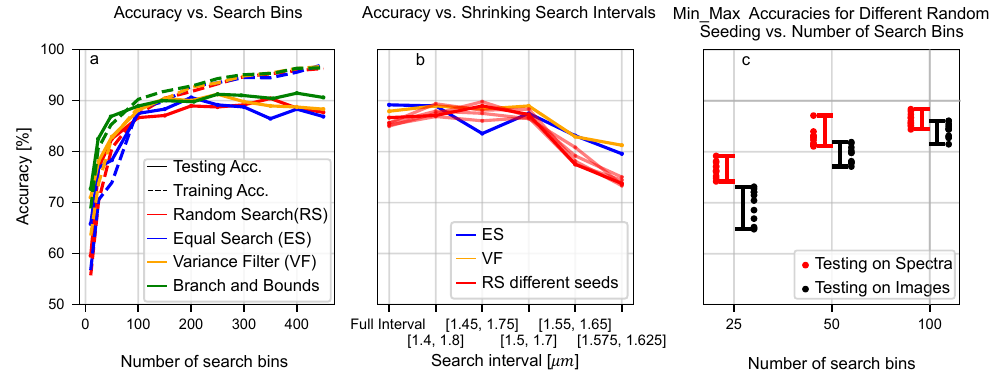}
    \caption{Performance of Random Search (RS) vs. optimized search methods on Handwritten MNIST images classification. (a) Testing and training accuracies achieved on MNIST vs. number of search bins for various search methods. (b) Test accuracies of different search methods vs. shrinking spectral search intervals. (c) RS test accuracies for different number of search bins when training on input images and output spectra vs. different random seeds.}
    \label{fig:placeholder}
\end{figure}

\subsection{Regression Regularization}
Linear regression is the training method used for reservoir computer \cite{rc_proposal}. However, reservoir states are often highly correlated because they are generated by a single dynamical system. Standard linear regression (ordinary least square) can fail or become unstable in these scenarios, whereas ridge regression stabilizes the solution by adding a penalty to the weights \cite{hoerl_ridge_1970}.\newline
When the number of samples and the dimensions of the sample does not match, the overdetermined and underdetermined conditions are fulfilled. These conditions may have no solution or infinitely many solutions. Tikhanov conditions with different normalizations help to solve this problem.\newline  
When L2-norm is implemented this type of regression is called ridge regression, it minimizes the regularized mean square error together with the Euclidean distance of the model parameters, hence less combinations of the available vectors. 
\begin{equation}
\mathcal{L}_{\text{Ridge}}(y_{\text{true}}, y_{\text{pred}}, u) = \mathcal{L}_{\text{MSE}}(y_{\text{true}}, y_{\text{pred}}) + \lambda \sum_{j=1}^{m} u_{j}^{2}
\end{equation}
The ridge factor $\lambda$ is incorporated in the closed-form matrix solution as shown previously in equation 1: 
\begin{equation*}
W = (X^TX + \lambda I)^{-1}X^Ty
\end{equation*}
When L1-norm is implemented, the type of regression is called LASSO, it minmizes the regularized mean square error together with the Manhattan distance of the model parameters.
\begin{equation}
\mathcal{L}_{\text{LASSO}}(y_{\text{true}}, y_{\text{pred}}, u) = \mathcal{L}_{\text{MSE}}(y_{\text{true}}, y_{\text{pred}}) + \lambda \sum_{j=1}^{m} |u_{j}|    
\end{equation}
Unlike Ridge regression, Lasso does not have a simple "closed-form" matrix solution.
The initial weight vector $\mathbf{w}_{init}$ is computed using the Moore-Penrose pseudoinverse of the Gram matrix:
\begin{equation}
    \mathbf{W}_{init} = (\mathbf{X}^T \mathbf{X})^{-1} \mathbf{X}^T \mathbf{y}
\end{equation}
Then we iteratively refine the weights, a scaling factor $\gamma$ is calculated for each dimension $i$ to provide non-uniform regularization:
\begin{equation}
    \gamma_i = \frac{1}{\sqrt{|\mathbf{w}_i|} + \epsilon}
\end{equation}
where $\epsilon = 10^{-6}$ is a small constant to ensure numerical stability.
The updated weight matrix $\mathbf{w}_{new}$ is then calculated by adding a weighted ridge penalty to the Gram matrix:
\begin{equation}
    \mathbf{W}_{new} = \left( \mathbf{X}^T \mathbf{X} + \lambda \mathbf{I} (\gamma \gamma^T) + \epsilon \mathbf{I} \right)^{-1} \mathbf{X}^T \mathbf{y}
\end{equation}
where $\lambda$ represents the ridge factor and $\mathbf{I}$ is the identity matrix, we iterate $k$ times updating the weight matrix every iteration.\newline
Both types of regularization prevent overfitting by either penalizing large parameter vaules (L2) or promotes sparsity (L1). For overdetermined systems, solutions minimize a loss function, often with regularization added. For underdetermined systems with infinitely many solutions, an additional criterion (minimum norm or regularization) selects a unique solution via the pseudoinverse. With regularization introduced, both cases use the same formulation. However, finding the regularization coefficient requires empirical tuning as shown in Figs. 7b–c, where we compare different L1 and L2 values, selecting L2 = 1 and L1 = 0.01 as the optimized solutions.\newline
Regularization impact is more pronunced when training the system on highly nonlinear task, therefore, we compare regularization effects on the spiral benchmark from our previous paper \cite{Saeed}. This classification task features 4 interleaved spirals (250 points each, maximum angular span $\theta_{max} = 4\pi)$; inputs are (x, y) coordinates per point as shown in Fig. 7a. We shuffled the input data while recording the output spectra to avoid chronological recording (class1–class2–class3, then class4), which boosts performance through seemingly-negligible intensity drift.\newline 
We compare linear regression without regularization, L2 regularization (ridge regression), and L1 regularization (LASSO) across two hidden layer filtering methods: Variance Filter (VF) and Equal Search (ES). In Fig. 7d, overfitting emerges around 250 bins achieving its peak accuracy of 90\% with ES at 200 bins before droping as the number of search bins increases. After incorporating regularization, overfitting becomes less detrimental: testing accuracy remains improves while training accuracy decreses by 5\%, yielding higher overall test accuracy of 93.5\% with 350 search bins and 92.5\% with 250 search bins using L2, L1, repectively (see Figs. 7e–f)\newline
Linear and ridge regression exhibit equivalent training complexity. In contrast, LASSO regression incurs higher computational costs, requiring up to 3× more time while using Variance Filter (VF) and between 5× and 10× more with Equal Search (ES), as shown in Figs. 7g–i and Table 2. This impact difference stems from VF’s output-oriented nature; applying LASSO regularization increases only the training complexity without influencing the readout selection process. Consequently, VF offers an efficient method for optimizing the L1 parameter, even when ES demonstrates superior final performance.

\begin{figure}[H]
    \centering
    \includegraphics[width=14cm]{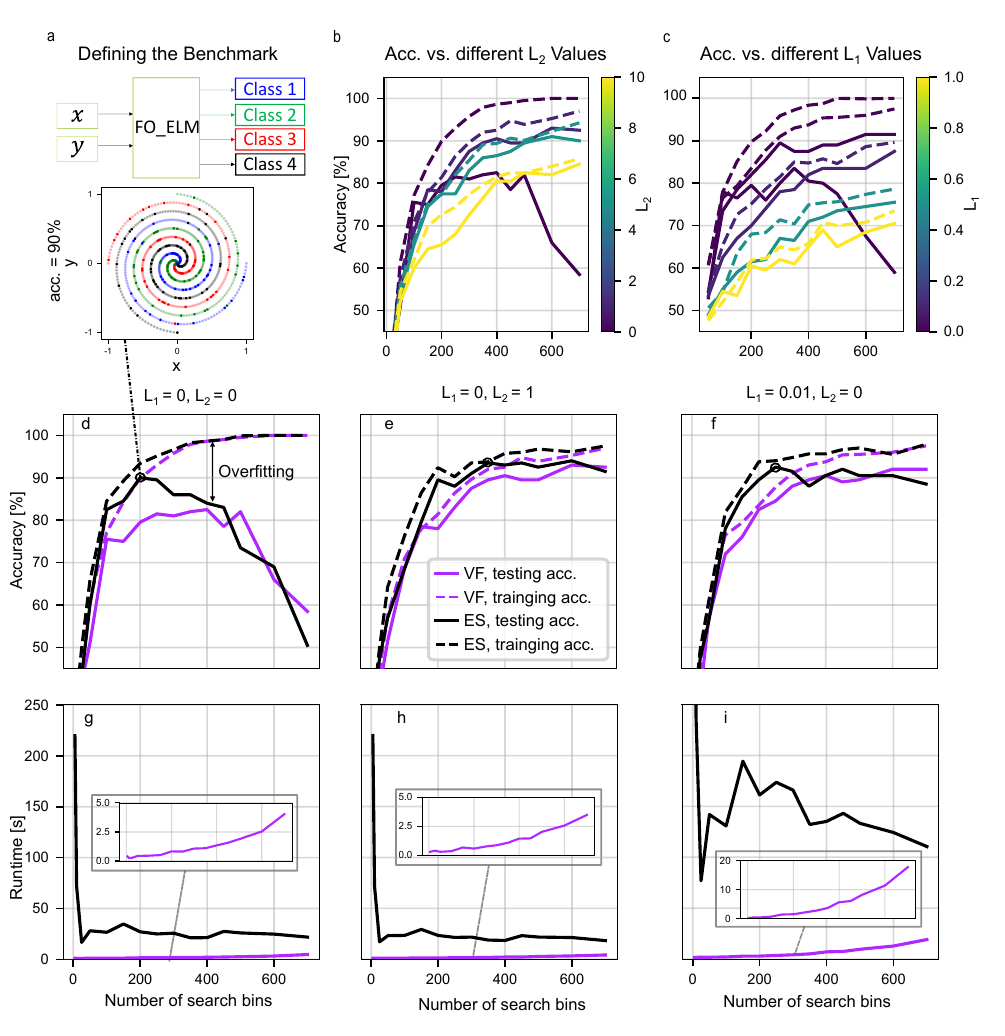}
    \caption{Impact of regularization on performance for a highly nonlinear task (4-spirals classification) using two search methods, Equal Search (ES) and Variance Filter (VF). (a) Defining the spiral benchmark: corresponding spectra to input coordinates (x,y) are classified into one of four spiral classes; bold points indicate classification results. (b–c) Training and testing accuracies vs. number of search bins for various L2 (b) and L1 (c) regularization strengths while using VF to select search bins. (d) Accuracies vs. number of search bins, highlighting significant overfitting in the absence of regularization for both search methods, VF and ES. (e–f) Accuracies vs. number of search bins for L2=1, L1=0 (e) and L2=0, L1=0.01 (f). (g–i) Search algorithm runtime vs. number of search bins for both search methods under various regularization schemes.}
    \label{fig:placeholder}
\end{figure}

\section{Discussion}
In this work, we addressed a key challenge in physical reservoir computing: the knapsack problem associated with reservoir hidden layer filtering. Specifically, we investigated strategies to mitigate overfitting and reduce computational cost through reservoir output sampling and regularization. We evaluated multiple search methods on MNIST handwritten digit classification and examined the impact of regularization techniques, particularly on the highly nonlinear task of spiral data classification.
We compared two loss-minimizing search methods: Equal Search (ES), which selects a combination of equidistant bins where the loss function is minimum, and Branch $\&$ Bound (BB), which iteratively removes a fixed number of bins, retaining the best possible combination at the end. These were benchmarked against an output-oriented statistical filtering approach, Variance Filter (VF), which selects bins with the highest standard deviation by analyzing output signals only. VF drastically reduces computational cost (less number of processes by order of ~1000) while achieving comparable or superior performance (testing accuracy on the spiral benchmark 90\% compared to 92\% achieved by ES, and highest achieved test accuracy on MNIST, 91.88\%), as shown in Tables 2 and 1. We further demonstrated that enforcing bin selection across the full output spectrum enhances the achieved accuracy (e.g. 9\% difference between prunned accuracies of VF best combination and VF global searh only, see table 1).
\begin{table}[H]
\centering
\caption{MNIST Accuracy Results for Different Pruning Methods}
\label{tab:mnist_results}
\resizebox{\textwidth}{!}{%
\begin{tabular}{lcccccccc}
\hline
\textbf{Maximum Accuracy (MNIST)}& ES & BB & VF (Global) & VF (Relative) & VF (Local) & VF (Best Comb.) & RS & No Pruning \\
\hline

Training Accuracy (\%) &
94.22 & 96.35 & 94.88 & 96.25 & 95.57 & 92.24 & 95.26 & 100.00 \\
Testing Accuracy (\%) &
91.25 & 91.46 & 87.92 & 86.25 & 90.00 & 91.88 & 90.42 & 36.88 \\
Bins &
300 & 400 & 350 & 450 & 350 & 200 & 350 & 2201 \\
\hline

\hline
\textbf{Pruned Accuracy (MNIST)}& ES & BB & VF (Global) & VF (Relative) & VF (Local) & VF (Best Comb.) & RS & No Pruning \\
\hline
Training Accuracy (\%) &
91.93 & 94.32 & 85.16 & 82.34 & 87.24 & 92.24 & 91.77 & -- \\
Testing Accuracy (\%) &
90.00 & 91.25 & 82.29 & 82.05 & 86.88 & 91.88 & 88.96 & -- \\
Bins &
200 & 250 & 150 & 200 & 100 & 200 & 200 & -- \\
\hline
\end{tabular}%
}
\end{table} 

\begin{table}[H]
\centering
\caption{Complexity Order and Number of Computations for Filtering and Training (in billions)\newline
N is the number of total bins (2700), n is the number of selected bins, m is the number of training samples (800), and k is the number of LASSO iterations (5), no gradient update rule was used inside the LASSO loop, so computations only scale up by the number of iterations}
\label{tab:complexity_computations}
\resizebox{\textwidth}{!}{%
\begin{tabular}{l c c c c c c c c c c}
\hline
\textbf{Regression Method} & 
\textbf{Filtering using ES} & 
\textbf{Filtering using VF} & 
\textbf{Training} & 
\textbf{No Filtering} & 
\multicolumn{2}{c}{\textbf{50 bins}} & 
\multicolumn{2}{c}{\textbf{200 bins}} & 
\multicolumn{2}{c}{\textbf{300 bins}} \\
\hline
& & & & \textbf{(2200 bins)} & \textbf{ES} & \textbf{VF} & \textbf{ES} & \textbf{VF} & \textbf{ES} & \textbf{VF} \\
\hline
LSQ/L2 & $O(N^2m^2/n)$ & $O(2Nm + nm^2)$ & $O(mn^2 + n^3)$ & 25.5 & 93.3 & 0.038 & 23.4 & 0.17 & 15.7 & 0.3 \\
L1 & $O((1+k)\frac{N^2m^2K}{n})$ & $O(2Nm + nm^2)$ & $O((1+k)(mn^2+n^3))$ & 153.1 & 556 & 0.05 & 140.4 & 0.37 & 94.3 & 0.8 \\
\hline
\end{tabular}%
}

\end{table}
This work also highlighted the trade-offs of random pruning. While it can achieve good performance (test accuracy on MNIST up to 89\%) due to its inherent property of generating a quasi-symmetric distribution, it lacks consistency. As we showed, test accuracy varies substantially depending on the employed random seed, and it degrades more rapidly compared to algorithmic pruning when the spectral search interval shrinks, highlighting the growing importance of informed readouts selection under constrained reservoir dimensions.
Two regularization methods were considered: L1 regularization (LASSO), which promotes sparsity in the solution but incurs additional computational cost (see Table 2), and L2 regularization (ridge regression), which suppresses large parameter values and improves robustness. Both LASSO and ridge regularization improved performance drastically on the spiral benchmark while having a negligible impact on MNIST. Efficient search strategies like VF are particularly advantageous for finding optimized regularization value.
Overall, these results emphasize the need to balance computational efficiency, pruning strategy, and regularization when designing practical physical reservoir computing systems.

\begin{backmatter}
\bmsection{Funding}
Carl-Zeiss Stiftung, Nexus program (P2021-05-025).
\bmsection{Acknowledgment}
We acknowledge that this work was made possible by funding from the Carl Zeiss Foundation
through the NEXUS program (project P2021-05-025).
\bmsection{Data availability}
The raw spiral benchmark data used are available on our Github page.  \url{https://github.com/Smart-Photonics-IPHT/09_Create-Data-Set_Create-Spiral-Data}
\bmsection{Conflict of interest}
The authors declare no conflict of interest.
\end{backmatter}

\bibliography{references.bib}

\end{document}